\documentclass[10pt,5p,times]{elsarticle}
\biboptions{sort&compress}
\usepackage{amssymb}
\usepackage{lipsum} 
\usepackage{slashed}
\usepackage{amsmath}
\usepackage{amsfonts}
\usepackage{xcolor}
\usepackage{placeins}
\usepackage{url}

\newcommand{\be}{\begin{equation}}
\newcommand{\ee}{\end{equation}}

\newcommand{\bea}{\begin{eqnarray}}
\newcommand{\eea}{\end{eqnarray}}

\newcommand{\dd}{\text{d}}
\newcommand{\Dmat}{{\bf D}}
\newcommand{\Gmat}{{\bf G}}
\newcommand{\Kmat}{{\bf K}}
\newcommand{\Wmat}{{\bf W}}

\journal{Physics Letters B}
\begin{document}

\begin{frontmatter}

\title{Accurate calculation of low energy scattering phase shifts of charged particles in a harmonic oscillator trap}

\author[heb]{Mirko Bagnarol}
\ead{mirko.bagnarol@mail.huji.ac.il}
\author[heb]{Nir Barnea}
\author[npi,char]{Mat\'u\v s Rojik}
\author[npi]{Martin Sch\"afer}
\ead{m.schafer@ujf.cas.cz}

\affiliation[heb]{organization={The Racah Institute of Physics}, addressline={The Hebrew University}, city={Jerusalem}, postcode={9190401}, country={Israel}}

\affiliation[npi]{organization={Nuclear Physics Institute of the Czech Academy of Sciences}, city={Rez}, postcode={25068}, country={Czech Republic}}

\affiliation[char]{organization={Faculty of Mathematics and Physics, Charles University}, city={Prague}, postcode={116 36}, country={Czech Republic} }

\begin{abstract}
Considering the elastic scattering of two charged particles, we present two methods for 
numerically solving the generalized Coulomb-corrected BERW formula with high accuracy 
across the entire energy spectrum. 
We illustrate these methods using $p-\alpha$ scattering, employing a phenomenological $p-\alpha$ short-range interaction. Our results reproduce the phase shifts computed with the Numerov method for all $l=0$ and $l=1$ channels. 
We also provide full access to the Python script used to obtain these results, which can be readily applied to a wide range of core-fragment scattering problems in nuclear and atomic physics.
\end{abstract}

\end{frontmatter}


\section{Introduction} \label{sec:intro}

The method of confining a scattering system dominated by short-range interactions in a harmonic oscillator (HO) trap was introduced by T. Busch, B.-G. Englert, K. Rzazewski, and M. Wilkens~\cite{BERW_original}. To study the interaction of two cold atoms, the authors related the scattering length in free space—where the particles interact through a short-range Dirac delta-like potential—to the bound state energies of the same system confined in a HO potential. This relationship became known as the Busch or BERW formula. The BERW formula was later generalized to connect trapped energies to the effective range expansion in free space, rather than just the scattering length. It has been proven for any short-ranged potential shape, not limited to Dirac deltas~\cite{Shea2009}. Finally, Suzuki et al. extended it to a scattering with an arbitrary orbital angular momentum $l$~\cite{Suzuki2009}.

While this approach has been successfully applied to several other cold atom studies (see~\cite{Mistakidis2023} for a review of recent advancements), it naturally fits in nuclear physics in systems at unitarity~\cite{Werner2006, Stetcu2010_v2, Hofmann2012, Tolle2012, Stetcu2013, Bazak2016} and for EFT predictions~\cite{Hammer2020}. In particular, in Pionless Effective Field Theory ($\slashed{\pi}$EFT) the nucleon-nucleon interaction is represented by regularized Dirac deltas and its derivatives, and therefore stands out as a particularly fitting theory for the BERW approach.
In $\slashed{\pi}$EFT the BERW formula has been applied with success to two~\cite{Luu2010, Stetcu2010, Li2021}, three~\cite{Rotureau2012}, four~\cite{Stetcu2007, Rotureau2010, Schafer2023} and five~\cite{Bagnarol2023} nucleon systems at LO and at NLO. 
Moreover, the systematic errors arising from the small distance matching of the wave functions in the BERW formula have also been addressed by X. Zhang~\cite{Zhang2020, Zhang2020_v2}.

Recently, P. Guo generalized the BERW formula to the case of two charged particles interacting via the Coulomb force in addition to a short-range interaction~\cite{Guo2021, Guo2021p2}. Even in this case, the relation between the free-space Coulomb phase shifts and the bound state energies of the trapped system can still be expressed in a closed formula. However, since the Hamiltonian with both the Coulomb and the HO potential cannot be solved analytically, the Coulomb and harmonic Green's function $G^{C,\omega}$ must be numerically calculated by solving the Dyson equation. The Dyson equation is a self-consistent equation that must be solved either by inversion, by iteration, or perturbative expansion. 

Recently, Zhang et al. \cite{Zhang2024} applied perturbative expansion to extract proton-proton $s$-wave phase shifts for the Minnesota potential \cite{Thompson1977} as the short-range interaction of choice. By including perturbative corrections up to eight order, they successfully reproduced the Minnesota phase shifts which were calculated using the $R$-matrix method. However, they observed convergence issues of the perturbative series once the bound state energies of the trapped system passed the position of the eigenenergies corresponding to non-interacting particles in the HO trap. This inconvenience hinders the extraction of phase shifts in certain energy regions.

In this paper, we propose two distinct numerical approaches to solve the Dyson equation for the Coulomb and Coulomb-plus-harmonic Green's functions, $G^{C}$ and $G^{C,\omega}$. The first method allows for an arbitrarily high number of iterations with controlled numerical stability, while the second method involves directly inverting the Dyson equation on a numerical grid. We demonstrate both methods using a phenomenological $N-\alpha$ potential~\cite{Ali1985} as our short-range interaction. The iterative approach faces similar convergence issues as the perturbative expansion. On the other hand, the matrix inversion offers a more robust tool on how to extract the scattering phase shifts at an arbitrary energy, except well-defined positions of discrete eigenenergies of the non-interacting HO-trapped particles.

The rest of the article is organized as follows. 
In Sec. \ref{sec:mod}, we show the relation between the free-space phase shifts and the bound state energies in a HO trap for a system of charged particles, and present the Dyson equations that yield $G^{C}$ and $G ^{C,\omega}$. 
In Sec. \ref{sec:meth}, we show our methods for solving numerically the Dyson equation, and discuss the stability with respect to the algorithms' parameters.
In Sec. \ref{sec:res}, we apply the methods to $p-\alpha$ scattering and discuss the results.
Summary and conclusions are presented in Sec. \ref{sec:con}.
In Appendix A access is given to the Python script we used to perform our calculation, along with proper documentation and a sample input.

\section{Theoretical background}\label{sec:mod}

The scattering phase shifts of two particles in an artificial trap satisfy the quantization condition~\cite{Guo2021, Guo2021p2, Zhang2024}:
\begin{equation}
\det \left[ {\rm cot}\left(\delta (E)\right) - \mathcal{F}^{\text{trap}} (E)    \right] = 0\label{eq:phsh_eq}
\end{equation}
where $\delta (E)$ is the scattering phase shift matrix diagonal in the orbital angular momentum $l$ and its third component $m$, while $\mathcal{F}^{\text{trap}}$ is a matrix depending on the geometry and dynamic properties of the trap.

Let us assume that we have a system of two particles with charges $q_1=Z_1 e$ and $q_2=Z_2 e$, which mutually interact through a short-range potential $V_s(r)$ and long-range Coulomb potential $V_C (r) = q_1 q_2/r$. By introducing a HO trapping potential $V_{\text{HO}}(r) = \frac{1}{2} \mu \omega ^2 r^2$ with an oscillator frequency $\omega$, we discretize the scattering continuum and give rise to a tower of infinite bound states with energies $E>0$. The Coulomb phase shifts $\delta^C(E)$ can then be extracted from the calculated bound state energies using Eq.~(\ref{eq:phsh_eq}) with a proper form of $\mathcal{F}^{\text{trap}}$. In this specific case~\cite{Guo2021, Guo2021p2, Zhang2024}:
\begin{align}
\mathcal{F}^{\text{trap}} (E) =& \left( 2 \mu k^{2l+1} C_l ^2 (\eta) \right)^{-1} \times \nonumber \\
& \times \lim _{r,r' \rightarrow 0} \frac{ \Re \left( G^C _l (r, r'; E) - G^{C,~\omega} _l (r,r'; E) \right) }{ (rr')^l} \label{eq:busch_coul}
\end{align}
where $r,\ r'$ stand for the inter particle distance, $\mu$ is the reduced mass, 
$k = \sqrt {2 \mu E}$ is the relative momentum,
\[ 
C_l (\eta) = \frac{2^l}{(2l+1)!} \left| \Gamma \left( l+1+i \eta \right) \right| e^{-\frac{\pi}{2} \eta},
\]
and $\eta = q_1 q_2 \mu / k$ is the dimensionless Sommerfeld parameter. $G^C _l (r, r')$ and $G^{C,~\omega} _l (r,r')$ are the Green functions for the Hamiltonian with only Coulomb and Coulomb plus HO potential, respectively. 

Both Green functions satisfy the Dyson equation, here reported in an integral form, as a second kind Fredholm equation:
\begin{align}
&G^C_l(r, r') = G_l ^{\text{free}} (r, r') + \int _0 ^{+\infty} \dd r''\  G_l ^{\text{free}}(r, r'') 
K (r'') G^C_l(r'', r')\nonumber \\
&G_l ^{C,~\omega}(r, r') = G_l ^{\omega} (r, r') + \int _0 ^{+\infty} \dd r''\  G_l ^{\omega}(r, r'') 
K (r'') G_l ^{C,~\omega}(r'', r')
\label{eq:dyson_general}
\end{align}
where $K (r'') = r''^2 V_C (r'') = q_1 q_2 r''$,
\begin{align}\label{eq:gfree0}
&G_l ^{\text{free}} (r, r'; E) = 
   -2 \mu i k~j_l\left(k r_<\right)~h ^{(+)}_l\left(k r_>\right)
\end{align}
is the free Green function, and
\begin{align}\label{eq:gtrapcoul0}
&G_l ^{\omega} (r, r'; E) = \nonumber \\
= &- \frac{(rr')^{-\frac{3}{2}}}{\omega} \frac{\Gamma \left( \frac{l}{2} + \frac{3}{4} - \frac{E}{2 \omega} \right)}{\Gamma \left( l + \frac{3}{2} \right)}  \mathcal{M}_{\frac{E}{2 \omega}, \frac{l}{2} + \frac{1}{4}} (\mu \omega r^2 _<) \mathcal{W}_{\frac{E}{2 \omega}, \frac{l}{2} + \frac{1}{4}} (\mu \omega r^2 _>) 
\end{align}
is the trap Green function for the relative Hamiltonian without the Coulomb potential. 
Here, $r_>$ ($r_<$) is the greater (lesser) between $(r,r')$, $j_l(x)$ and $h ^{(+)}_l(x)$ are the regular spherical Bessel and Hankel functions, and $\mathcal{M}_{\kappa, \nu} (x)$ and $\mathcal{W}_{\kappa, \nu} (x)$ are the Whittaker functions. All in all, we can calculate the Coulomb phase shift $\delta^C(E)$ of any system through Eqs.~\eqref{eq:phsh_eq}~and~\eqref{eq:busch_coul} by first solving Eqs.~\eqref{eq:dyson_general} for $G^{C} (r, r'; E)$ and $G ^{C,\omega} (r, r'; E)$.

\section{Method} \label{sec:meth}

We consider two methods to calculate $\mathcal{F}^{\text{trap}}$ through Eq.~\eqref{eq:busch_coul}. They differ in the implemented numerical techniques applied to obtain $G_l^C(r,r')$ and $G_l^{C,\omega}(r,r')$ as solutions of the Dyson equations~\eqref{eq:dyson_general}. The Green functions are calculated on a numerical grid for $r$ and $r'$ by fixing $r_{\text{min}}$ as the starting point and $r_{\text{max}} = f \cdot b$ as the endpoint. Here, $b = \sqrt{\frac{1}{\mu \omega }}$ is the length of the harmonic oscillator trap, and $f$ is a numerical factor. We divide the interval $[r_{\text{min}}, r_{\text{max}}]$ in $N$ grid points with progressively rising steps, such that
\[
r_n = r_{n-1} + h q^n
\]
where $q$ is the incremental ratio and $h$ is the step, fixed such that $r_N = r_{\text{max}}$. If $q = 1$, the $N$ points are equidistant.  The limit of Eq. \eqref{eq:busch_coul} is approximated as 
\begin{equation}
\frac{ \Re \left( G^C _l (r_{\text{min}}, r_{\text{min}}, E) - G^{C,~\omega} _l (r_{\text{min}},r_{\text{min}}, E) \right) }{ (r_{\text{min}})^{2l}}. \label{eq:numeric_lim}
\end{equation}

On the grid $\{r_n\}$ the Green functions are simply represented as the matrices
\begin{equation}
  G_{ij}=G(r_i,r_j)
\end{equation}
and the Dyson equations become simple matrix equations
\begin{align}\label{Dyson_matrix_form}
\Gmat_l^C ~~ &= \Gmat_l^{\text{free}} + \Gmat_l^{\text{free}}~\Dmat\Kmat\Gmat_l^C,
\cr 
\Gmat_l^{C,\omega} &= \Gmat_l^{\omega} ~+ \Gmat_l^{\omega}~\Dmat\Kmat\Gmat_l^{C,\omega},
\end{align}
where bold letter denote a matrix, $K_{ij}=\delta_{ij}K(r_i)$ and $\Dmat$ is the diagonal matrix of integration weights. 

\subsection{Successive approximation method with stabilization}

The first numerical approach to solving Eqs. \eqref{eq:dyson_general} is a two-component process: (\emph{i}) it self-consistently loops Dyson equations until a convergence criterion is achieved, and (\emph{ii}) it stabilizes the function numerically with a feedback mixing parameter $0 < \epsilon \leq 1$.

The initial $(m=0)$ step starts by defining $\Gmat_l^{C}(0) ~= \Gmat_l^{\text{free}}$ and $\Gmat_l^{C,\omega}(0) = \Gmat_l^{\omega}$ as the free and HO Green functions 
(matrices) in the absence of the Coulomb interaction (Eqs.~\ref{eq:gtrapcoul0}), respectively. The first two steps ($m=1$ and $m=2$) are:
\begin{align}
Iteration~~~~~~~~~~~~~&\cr
  \Gmat_l^{C}(1) ~&= \Gmat_l^{\text{free}} + \Gmat_l^{\text{free}}~\Dmat\Kmat
                   \Gmat_l^{C}(0),
  \cr 
  \Gmat_l^{C,\omega}(1) &= \Gmat_l^{\omega} ~+ \Gmat_l^{\omega}~\Dmat\Kmat
  \Gmat_l^{C,\omega}(0),
\cr
Feedback~mixing&\nonumber\\
    \Gmat_l^{C}(2)~~~ &=  \epsilon ~\Gmat_l^{C}(1)\ +  
                         (1 - \epsilon)\Gmat_l^{C}(0)
    \cr
    \Gmat_l^{C,\omega}(2) &=  \epsilon \Gmat_l^{C,\omega}(1) + 
         (1 - \epsilon)\Gmat_l^{C,\omega}(0)
    \label{eq:first_step}
\end{align}
For the $(2m+1)$th and $(2m+2)$th steps one should replace $(0)~\rightarrow~(2m)$, $(1)~\rightarrow~(2m+1)$ and $(2)~\rightarrow~(2m+2)$. We establish that the convergence has been reached when
\begin{equation}
    \left| \text{Trace}\Big(\Dmat\left[ \Gmat(2m+1) - \Gmat(2m) 
      \right]\Big)\right| < \Delta \nonumber
\end{equation}
for both Green functions $G_{l}^C$ and $G_{l}^{C,\omega}$. In our calculations, the numerical parameter $\Delta$ is fixed to the value $\Delta= 10^{-8}$. However, the calculated phase shifts change by less than $1\%$ when $\Delta \leq 10^{-3}$. While checking for the convergence, we ignore the off-diagonal points in calculated Green functions: if the limit in Eq. \eqref{eq:busch_coul} exists, it can be taken along the $r=r'$ diagonal. Once the diagonal of both Green's functions is converged, we can apply Eq. \eqref{eq:busch_coul} without further convergence demands elsewhere.

\begin{figure*}
    \centering %
    \begin{tabular}{cc}
    \includegraphics[width = \columnwidth]{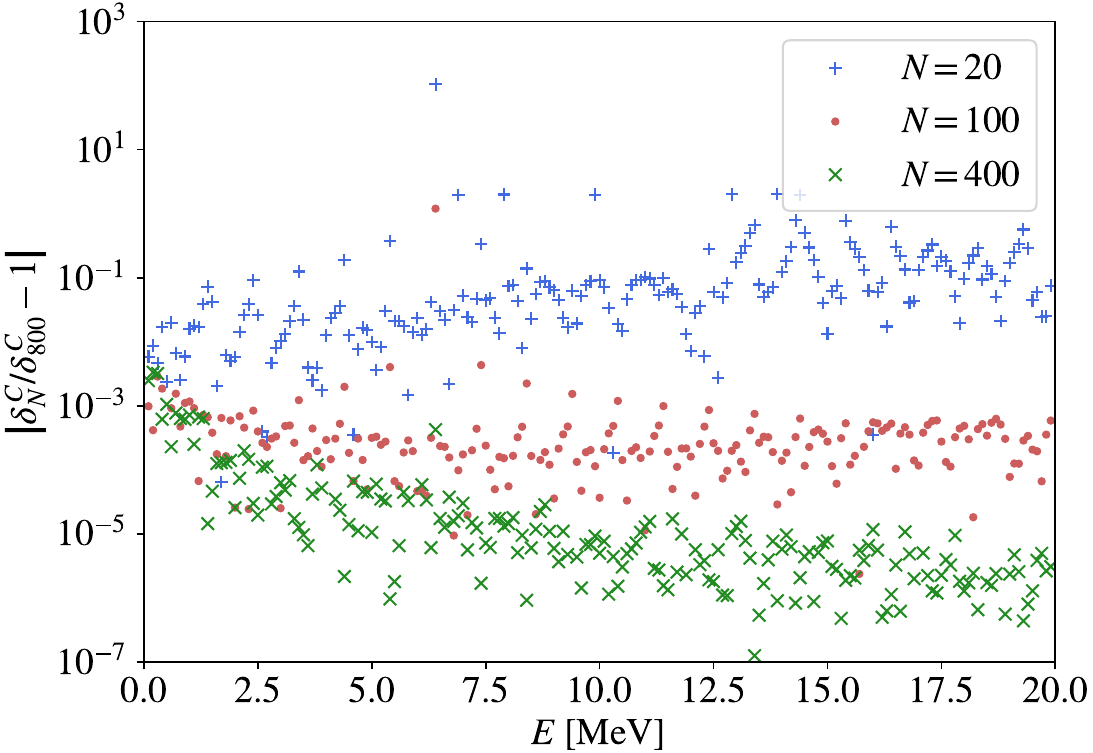} & 
    \includegraphics[width = \columnwidth]{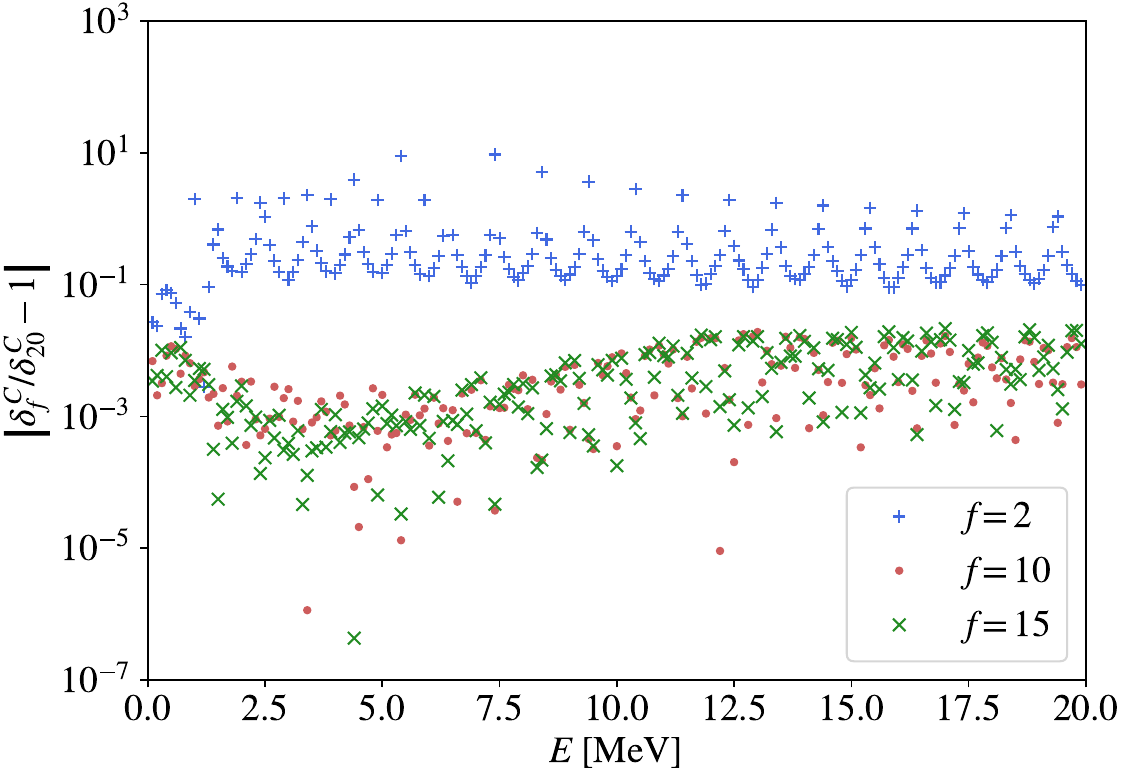}\\
    \end{tabular}
    \caption{Sensitivity of the $p-\alpha$ phase shifts, calculated using the direct method and Eqs.~\eqref{eq:phsh_eq}~and~\eqref{eq:busch_coul}, at different relative energies $E$ to the numerical parameters $N$ and $f$. The HO trap frequency is $\omega=0.5$~MeV. Left panel: relative variation of $\delta^C(E)$ when the number of grid points changes between $20 \leq N \leq 400$ with respect to $N=800$, while the endpoint is fixed at $f=10$. Right panel: same as the left panel but with a fixed number of grid points $N=400$ and with varying $2 \leq f \leq 15$, while the control phase shifts are computed with $f=20$.}
    \label{fig:numerical_study}
\end{figure*}

\subsection{Direct method}
A more direct approach to solve the Dyson equation, applicable when the problem size 
is not too large, is to solve it through direct
inversion.
Given the matrix form of the Dyson equations \eqref{Dyson_matrix_form},
the Green functions are obtained through matrix inversion
\begin{align}
{\bf G}_l^C &= \left({\bf 1} - {\bf W}_l^{\text{free}}\right)^{-1} ~ {\bf G}_l^{\text{free}},\nonumber\\
{\bf G}_l^{C,\omega} &= \left({\bf 1} - {\bf W}_l^{\omega}\right)^{-1} ~ {\bf G}_l^{\omega},
\end{align}
where
\begin{align}
  \Wmat_l^{\text{free}} &= \Gmat_l^{\text{free}} \Dmat \Kmat~,
  \cr
  \Wmat_l^{\omega}   &= \Gmat_l^\omega \Dmat\Kmat~.
\end{align}
\subsection{Numerical considerations}
In the listed approaches, $r_\text{min}$ represents a numerical parameter that should be selected to balance the accuracy of limit approximation in Eq.~\eqref{eq:numeric_lim} and numerical precision. By decreasing $r_\text{min}$, both the numerator and denominator in the limit acquire small values. However, the $G^C _l (r, r')$ and $G^{C,~\omega} _l(r,r')$ separately diverge for $r,r' \rightarrow 0$. This translates into relatively large but similar Green function values in the numerator at the given $r_\text{min}$. Eventually, by selecting too small $r_\text{min}$ value, one encounters the numerical issue with the floating point numbers subtraction and a loss of numerical precision.

The Gamma function $\Gamma \left( l/2+3/4-E/2 \omega \right)$ in $G^{\omega}_l(r,r',E)$, Eq.~\eqref{eq:gtrapcoul0}, has poles at $E/\omega=l+3/2+2n$, where $n=0,1,2,...$. In the very vicinity of these points, the iterative solution in the successive approximation method does not converge, and in the direct method, the ${\bf 1} - {\bf W}_l^{\omega}$ matrix can not be inverted to obtain ${\bf G}_l^{C,\omega}$ solution.


\section{Results} \label{sec:res}

To demonstrate the applicability of our numerical methods, we consider a two-body $p-\alpha$ scattering, where the $\alpha$ nucleus is treated as a point-like particle. 
In our calculations we have set the reduced mass to be $\mu = 4/5 m_N$, where $m_N$ is the mass of the nucleon, the mass parameter $(\hbar c)^2/m_N=41.47~\rm{MeV\ fm^2}$, $e^2 = 1.44~\rm{MeV\ fm}$, $r_{\text{min}} = 10^{-3}$~fm, and $q=1.025$. If $r_{\text{min}} \leq 10^{-4}$ fm, the ratio in Eq.~\eqref{eq:numeric_lim} becomes numerically unstable and our results rapidly deteriorate, while for $10^{-3}$ fm $\leq r_{\text{min}} \leq 10^{-1}$ fm the extracted phase shifts are numerically stable and vary by less than $1\%$. We find that once the successive approximation method converges, the corresponding phase shift values are mostly identical to those extracted using the direct method. Consequently, we will predominantly present the results obtained using the direct method. Both approaches will be compared in the last part of this section, where we will also comment on the convergence issues of the successive approximation method.

In the left panel of Fig.~\ref{fig:numerical_study}, we show stabilization of $p$-wave phase shifts at different energies $E$, calculated using the direct method and Eq.~\eqref{eq:phsh_eq}, with an increasing number of grid points $N$ and $\omega =0.5$~MeV. There is less than 1\% change in the calculated phase shift values by increasing $N=400$ to $N=800$. The stabilization with increasing parameter $f$, which defines the endpoint $r_\text{max}$ in a numerical integration, is demonstrated in the right panel of Fig.~\ref{fig:numerical_study}. If $f\geq 10$, the accuracy of extracted phase shifts remains below $1\%$, while $f=2$ appears too low.

\begin{figure}
    \centering %
    \includegraphics[width = \columnwidth]{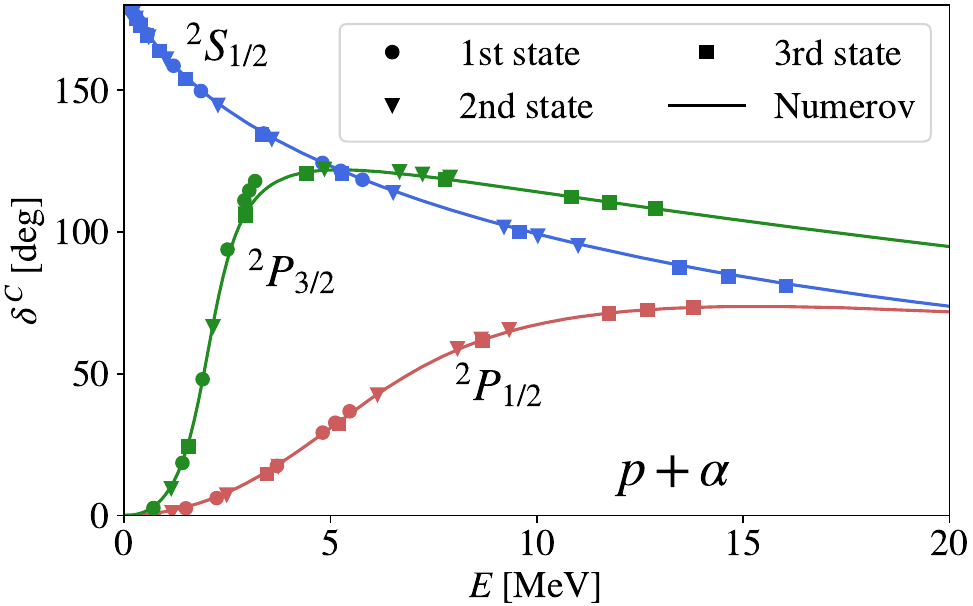}
    \caption{The $p-\alpha$ $s$-wave ${}^2 S_{1/2}$ (blue) and $p$-wave ${}^2P_{1/2}$ (red), ${}^2P_{3/2}$ (green) phase shifts, for the potential in Eq.~\eqref{eq:pa_pot}, displayed as a function of the relative energy $E$. The circles, triangles, and squares represent the phase-shift values extracted using the direct method from the HO trap energies corresponding to the discretized continuum's first, second, and third lowest state, respectively. The solid line denotes phase shifts calculated in free space using the Numerov method and matching at large distances the scattering wave function to the Coulomb functions (no harmonic oscillator trap is involved).}    
    \label{fig:pa_phsh}
\end{figure}

We test our methods by calculating scattering phase shifts of a simple $p-\alpha$ phenomenological potential~\cite{Ali1985} 
\begin{equation}
V_{p-\alpha}(r) = \left( 1 + \beta\ {\pmb l} \cdot {\pmb \sigma} \right) V(r),
\label{eq:pa_pot}
\end{equation}
where $V(r)=V_0$ for $r<a$, $V(r)=0$ for $r>a$, and the potential parameters $V_0=-33.0$~MeV, $a=2.55$~fm, $\beta=0.103$ were obtained from the fit to the experimental $p-\alpha$ phase shifts in $^2S_{1/2}$, $^2P_{1/2}$, and $^2P_{3/2}$ partial waves. We solve the two-body Schr\"{o}dinger equation with the Hamiltonian
\begin{equation}
H = T_k + V_{p-\alpha}(r) + \frac{2 e^2}{r} + \frac{1}{2} \mu \omega ^2 r^2,
\end{equation}
where $T_k$ is the relative kinetic energy, for the HO frequency values $0.015 \text{ MeV} \leq \omega \leq 2.35 \text{ MeV}$, corresponding to the HO trap length values $4.7 \text{ fm} \leq b \leq 60 \text{ fm}$. We get the energies of the three lowest bound states of the discretized continuum $E>0$. Finally, the phase shifts $\delta^C(E)$ are extracted for each couple $(E, \omega)$ using the direct method outlined in Sec.~3, with $N=400$ grid points and $f=10$.

In Fig.~\ref{fig:pa_phsh}, we compare the $p-\alpha$ phase shifts extracted from the three lowest HO trap energies to the phase shifts obtained using the Numerov algorithm and matching the calculated scattering wave function to the long-range Coulomb asymptotic. For the HO trap frequencies $\omega \lesssim 0.7$~MeV (corresponding to the HO trap length $b\gtrsim 6$~fm), the extracted $p-\alpha$ phase shifts differ by less than 1\% from the results of the Numerov free-space calculations (solid lines). For larger $\omega$, we observe a systematic error introduced by the not sufficiently separated $p-\alpha$ interaction range and size of the HO trap. This is especially visible in the figure for the $^2P_{3/2}$ phase shifts around $E=2.5$~MeV, which are extracted using the discretized continuum's first (lowest) bound state (green dots). Here, the phase shifts start to significantly deviate from the free-space results. This systematic error can be either suppressed by using sufficiently small HO trap frequencies $\omega$ or systematically removed by subtracting residual HO trap dependencies as proposed for the neutral particle scattering \cite{Zhang2020}.

\begin{figure}
    \centering %
    \includegraphics[width = \columnwidth]{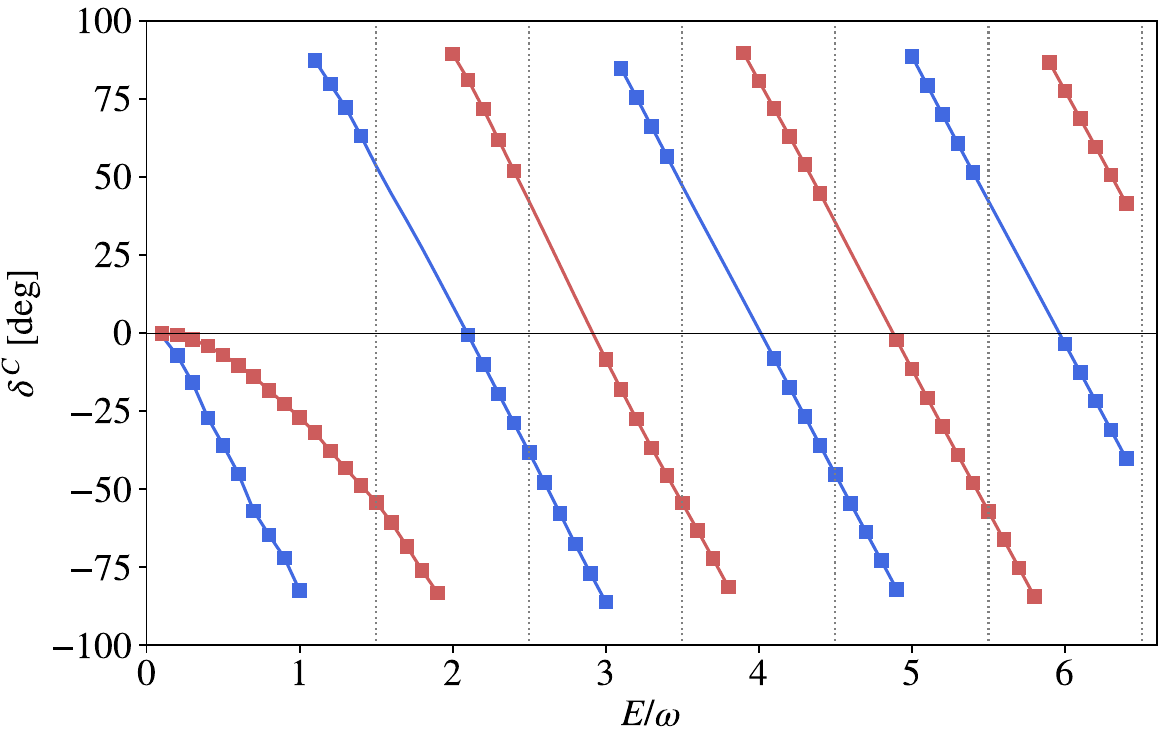}
    \caption{The $s$-wave (blue) and $p$-wave (red) $p-\alpha$ phase shifts are displayed as a function of the relative energy $E$ divided by the HO frequency $\omega$. The phase shifts are calculated using $\omega=0.5$~MeV, Eqs.~\eqref{eq:phsh_eq} and \eqref{eq:busch_coul}, and the Coulomb and HO plus Coulomb Green functions obtained using either direct (solid lines) or successive approximation (squares) method. The vertical grey dotted lines show the position of the poles in $G^{\omega}_l(r,r',E)$ Green function, Eq.~\eqref{eq:gtrapcoul0}, at $E/\omega =l + 3/2 + 2n$ for $l=0,~1$ and $n\in \mathbb{N}$.}    
    \label{fig:e_span}
\end{figure}

We demonstrate in Fig.~\ref{fig:e_span} that the $s$- and $p$-wave $p-\alpha$ phase shifts, which are obtained by solving Eqs.~\eqref{eq:phsh_eq}~and~\eqref{eq:busch_coul} using the direct or successive approximation method, are in agreement. However, we notice that the successive approximation method is not always able to find the converged $G_{l}^C(r,r')$ or $G_{l}^{C,~\omega}(r,r')$ solution. For $G_{l}^C(r,r')$ and the $p-\alpha$ system, this happens at very low energies where the Coulomb interaction dominates. For $G_{l}^{C,~\omega}(r,r')$, we observe convergence issues for positive phase shift after the $E/\omega$ value passes the pole position introduced in the corresponding Dyson equation by the Gamma function in $G_{l}^{\omega}(r,r')$, Eq.~\eqref{eq:gtrapcoul0}. While approaching this region, the number of iterations necessary to obtain an accurate $G_l^{C,~\omega}$ solution dramatically increases, and by passing the pole position the iterative series diverges. This is demonstrated by missing squares for positive phase shifts in Fig.~\ref{fig:e_span}. The same convergence issue was reported in
Ref.~\cite{Zhang2024} for the perturbative approach.

\section{Conclusions} \label{sec:con}

Using the HO trap technique, we proposed an efficient way to extract free-space Coulomb phase shifts of two-charged-particle elastic scattering with Coulomb and short-range interaction. We applied direct and successive approximation methods to numerically solve the Dyson equation for Coulomb and Coulomb plus HO trap Green functions. These solutions were then used to connect the HO trap's energy spectrum of two charged particles to the corresponding free-space Coulomb phase shifts. 

We tested our methods by calculating $p-\alpha$ Coulomb phase shifts. We observed that once the successive method converges, both the direct and the successive approximation methods yield almost identical results that change by less than 1\% with the proper selection of the numerical grid. The integral part of this letter is the Python script, briefly introduced in Appendix A. The script employs our methods and can be directly used to calculate Coulomb phase shifts from HO trap energies.

We emphasize that the developed tools are general and can be directly applied to study the elastic scattering of any charged projectile and target. This does not necessarily involve only particle-particle scattering but also the scattering of two composite objects. For example, in Refs.~\cite{Schafer2023,Bagnarol2023}, the HO trap technique without Coulomb was successfully applied to calculate phase shifts in few-body systems.

Of particular importance is the application of the presented methods in HO trap techniques involving coupled-channel and inelastic scattering \cite{Guo2022, Zhang2024_2}. This will be especially relevant in future HO trap \emph{ab initio} studies of few- and many-body nuclear reactions, where multiple incoming and outgoing channels of charged nuclei must often be considered to obtain a realistic description of scattering dynamics.

\section*{Acknowledgement}
The work of M. B. and N. B. was supported by the Israel Science Foundation grant 1086/21. The work of M. R. and M. S. was supported by the Czech Science Foundation GA\v{C}R grant 22-14497S. 

\section*{Appendix A}

The Python script is available at \cite{script} and is released under the GNU General Public Licence \cite{gpl}. The script uses NumPy~\cite{2020NumPy-Array}, SciPy~\cite{scypy}, and MPMath~\cite{mpmath} libraries.

The input is given either from a \texttt{json} file or in a shell-style series of inputs. A detailed description of each input if found by typing \texttt{python3~busch\_coulomb.py~-h} in the shell input or in the \texttt{README} documentation file. The dependence on the most important input parameters are discussed throughout the paper, in particular in Fig. \ref{fig:numerical_study}. The method can be chosen between successive approximation and direct inversion.

The script writes in output the phase shift and the effective range expansion for each energy $E$ and $\omega$ pair in input. We remind that the effective range expansion in presence of Coulomb interaction is~\cite{Arndt1973}:
\begin{equation*}
k^{2l + 1} C_l ^2 (\eta) \left[ \cot \delta _l + \frac{2 \eta\  h(\eta)}{C_0 ^2 (\eta)} \right] = - \frac{1}{a_l} + \frac{1}{2} r_l k^2 - \frac{1}{4} P_l ^{(0)} k^4 + \dots, 
\end{equation*}
where $a_l$ is the scattering length, $r_l$ is the effecting range, $P_l ^{(n)}$ are the shape parameters, 
\[
h(\eta) = \frac{\psi(1 + i \eta) + \psi(1 - i \eta)}{2} - \ln \eta 
\]
and $\psi(z) = \frac{\dd}{\dd z} \ln \Gamma(z)$ being the logarithmic derivative of the Gamma function. The extraction of the scattering parameters $a_l$ and $r_l$ is supposed to be performed separately by the user.


\FloatBarrier
\bibliographystyle{unsrt}
\bibliography{refs}


\end{document}